\documentclass[aps,preprint,preprintnumbers,epsf,amsmath,amssymb,nofootinbib]{revtex4}
\usepackage{amssymb}
\usepackage{amsmath}
\usepackage{bm}
\usepackage{graphicx}
\usepackage{epsfig}
\draft
\begin{document}
\title{Natural exit of fresh inflation to a radiation dominated universe}
\author{$^{1, 2}$Mauricio
Bellini\footnote{mbellini@mdp.edu.ar}}
\address{$^{1}$ Departamento de F\'{\i}sica, Facultad de Ciencias Exactas y
Naturales, \\
Universidad Nacional de Mar del Plata, Funes 3350, (7600) Mar del
Plata, Argentina. \\ \\
$^{2}$ Instituto de Investigaciones F\'{\i}sicas de Mar del Plata
(IFIMAR), Consejo Nacional de Investigaciones Cient\'{\i}ficas y
T\'ecnicas (CONICET), Argentina.}
%%%%%%%%%%%%%%%%%%%%%%%%%%%%%%%%%%%%%%%%%%%%%%%%%%%%%%%%%%%%%%%%%%%%%%%%%%%%%%%%%%%%%%%%%%%%%%%%%%%%%%%%%%%%%%%%%%%%%%%%%%%%
\begin{abstract}
We examine the transition from a fresh inflationary scenario to a
radiation dominated universe when the inflaton field is coupled to
gravity. We show that this transition is very natural when this
coupling is $\xi=1/4$ at the end of fresh inflation. This is a
very important result that shows as fresh inflation has a natural
exit to the radiation dominated epoch.
\end{abstract}
%%%%%%%%%%%%%%%%%%%%%%%%%%%%%%%%%%%%%%%%%%%%%%%%%%%%%%%%%%%%%%%%%%%%%%%%%%%%%%%%%%%%%%%%%%%%%%%%%%%%%%%%%%%%%%%%%%%%%%%%%%%%
\maketitle

\section{Introduction and Motivation}

Inflationary cosmology\cite{Guth,Al} is the most strong candidate
to explain the isotropic, homogeneous and flatness nature of the
universe on cosmological scales. This fact is supported by
experimental evidence\cite{smoot}. In particular, the model of
fresh inflation\cite{fresh} attempts to build a bridge between the
standard and warm inflationary models and hence can be viewed as
an unification of both, chaotic and warm inflation scenarios. In
this model the universe begins from an unstable primordial matter
field perturbation with energy density nearly $M^4_p$ and chaotic
initial conditions to later describing a second order phase
transition with heating and particles production. Hence, radiation
energy density grows during fresh inflation because the Yukawa
interaction between the inflaton field and other fields in a
thermal bath lead to dissipation which is responsible for the slow
rolling of the inflaton field though the minimum energetic
configuration. Hence, slow-roll conditions are physically
justified and there are not a requirement of a nearly flat
potential in fresh inflation. As a consequence of the strong
dissipation produced by the Yukawa interaction $\Gamma \gg H$),
there is no oscillation of the inflaton field around the minimum
of the effective potential\cite{berera}.

A very interesting issue to study is the natural exit of fresh
inflation to a radiation dominated epoch. In this work we examine
this topic in the framework of a scalar field which is coupled to
gravity though a coupling $\xi$. In this context one can see that
for $\xi=1/4$ radiation dominance becomes possible when the scalar
field $\phi$ approaches to the minimum energetic configuration.

\subsection{Nonminimal coupling in fresh inflation revisited}

The system is considered though a Lagrangian density of an
inflaton field which is coupled to the scalar curvature $R$, plus
a self-interaction Lagrangian ${\cal L}_{int}=-g^2 \,\phi^4$,
which describes the self-interaction of the inflaton
field\cite{coupling}
\begin{equation}
{\cal L} = -\sqrt{-g} \left[ \frac{R}{16\pi G} - \frac{1}{2}
\partial^{\mu} \phi \partial_{\mu}  \phi + V(\phi) + \frac{1}{2}
\xi R \phi^2 + {\cal L}_{int}\right],
\end{equation}
where $G=M^2_p$ is the gravitational constant, $M_p = 1.2 \times
10^{19}\ {\rm GeV}$ is the Planckian mass. During fresh inflation
$\xi \ll 1/4$, but at the end of fresh inflation it can hold to
values close to $\xi=1/4$. In this paper we are interested to
study this particular case. We shall consider a 4D
Friedmann-Robertson-Walker (FRW) metric with a line element given
by $ds^2 = dt^2 - a^2(t) dr^2$, such that $dr^2 = dx^2+dy^2+dz^2$.
The diagonal Einstein equations are given by
\begin{eqnarray}
3 H^2 & = & 8\pi G \left[ \left(\frac{1}{2} - 2\xi\right)
\dot\phi^2 +
V(\phi) + \rho_r \right], \label{3}\\
3 H^2 + 2 \dot H & = & -8\pi G \left[ \left(\frac{1}{2} -
2\xi\right) \dot\phi^2 - V(\phi) + \frac{\rho_r}{3}\right],
\label{4}
\end{eqnarray}
where $H=\dot a/a$ is the Hubble parameter and $a$ is the scale
factor of the universe which is considered with zero spatial
curvature. If $\delta=\dot\rho_r + 3 F H \rho_r=\Gamma(\theta)
\dot\phi^2$\footnote{Other phenomenological interaction terms as
$\delta\propto \phi^2\dot\phi^2$\cite{14} or $\delta\propto
\phi^{5-2d}\dot\phi^d$\cite{17}, has been proposed in the
literature.} describes a Yukawa interaction between the inflaton
field and the thermal bath, the equations of motion for $\phi$ and
$\rho_r$, hold
\begin{eqnarray}
&& \ddot\phi + 3 H \dot\phi + V'(\phi) + \xi R \phi +
\frac{\delta}{\dot\phi} =0, \\
&& \dot\rho_r + 3 F H \rho_r -\delta =0.
\end{eqnarray}
The parameter $F=1+\omega$, such that $\omega=p_t/\rho_t$ give us
the equation of state of the universe, or
\begin{equation}\label{7}
F= -\frac{2 \dot H}{3 H^2} = \frac{(1-4\xi) \dot\phi^2+
\frac{4}{3} \rho_r}{\rho_r + \left(\frac{1}{2} - 2\xi\right)
\dot\phi^2 + V} >0,
\end{equation}
with a scalar curvature $R= 12 H^2+6\dot{H}$. In the early stages
of inflation this parameter is very small: $F\ll 1$, but at the
end of inflation it reaches values close to $F=4/3$, so that, for
couplings close $\xi \simeq 1/4$, inflation ends in a radiation
dominated era with $a\sim t^{1/2}$. From the inequalities in eq.
(\ref{7}) we obtain
\begin{eqnarray}
&& \dot\phi^2 \left[(2-F)\left(\frac{1}{2} + 2\xi\right)\right] +
\rho_r
\left(\frac{4}{3} - F\right) - F V(\phi)=0, \label{8} \\
&& H = \frac{2}{3 \int F dt}, \label{9}
\end{eqnarray}
where, because $\dot H = H' \dot\phi$, one obtains
\begin{equation}\label{10}
\dot\phi = -\frac{3 H^2}{2 H'} F.
\end{equation}
By replacing eq. (\ref{10}) in eq. (\ref{8}) and later in eq.
(\ref{4}), we obtain respectively the eduations for the radiation
energy density and the potential $V(\phi)$
\begin{eqnarray}
&& \rho_r = \left(\frac{3F}{4-3F}\right) V(\phi) - \frac{27}{4}
\left(\frac{H^2}{H'}\right)^2 F^2 \left[ \frac{(2-F)}{(4-3F)}
\left(\frac{1}{2} + 2\xi\right)\right], \\
V(\phi)& = & \frac{3(1-8\pi \xi G \phi^2)}{8\pi G} \left[
\left(\frac{4-3F}{4}\right) H^2 + \frac{3\pi}{2} \frac{G
F^2}{(1-8\pi \xi G \phi^2)} \left(\frac{H^2}{H'}\right)^2
(1+8\xi)\right]. \end{eqnarray}

\subsection{The model}

Fresh inflation is described by a global group $O(n)$ that involve
a single $n$-vector multiplet of scalar fields
$\phi_i$\cite{fresh}, where the effective inflaton field is
describes by the norm of their scalar components:
$(\phi^i\phi_i)^{1/2}$. The effective potential is described by
$V_{eff}(\phi,\theta)  = V(\phi) + \rho_r(\phi,\theta)$, such that
\begin{equation}
V_{eff}(\phi,\theta) = \frac{{\cal M}^2(\theta)}{2} \phi^2 +
\frac{\lambda^2}{4} \phi^4,
\end{equation}
where the effective squared mass ${\cal M}^2(\theta) = {\cal
M}^2(0) + {(n+2)\over 12 } \lambda^2 \theta^2$, ${\cal M}^2(0)
\simeq 10^{-12}\, {\rm M^2_p}$ and $n$ is the number of created
particles due to the interaction of $\phi$ with the particles in
the thermal bath
\begin{equation}
(n+2) = \frac{2\pi^2}{5 \lambda^2} g_{eff}
\frac{\theta^2}{\phi^2}.
\end{equation}
During the fresh inflationary epoch $\dot\rho_r \simeq \delta >0$,
so that $\rho_r$ increases with the time. Therefore, the number of
particles created is increased dramatically because
$\phi(t)=1/(\lambda t)$, but the temperature $\theta \sim
\rho_r^{1/4}$ increases with time. This means that the ratio
$\frac{\theta^2}{\phi^2}$ will be increased as fresh inflation
evolves.

\subsection{Dynamics of the fluctuations $\delta\phi(\vec{x},t)$}

The dynamics of the background scalar field $\phi(t)$ and the
fluctuations $\delta\phi(\vec{x},t)$ are described respectively by
the equations
\begin{eqnarray}
&& \ddot\phi + (3H + \Gamma) \dot\phi + \xi R \phi + V'(\phi)=0,
\\
&& \ddot{\delta\phi} -\frac{1}{a^2} \nabla^2 \delta\phi + (3H
+\Gamma) \dot{\delta\phi}  + \left[ \xi R + V''(\phi)\right]
\delta\phi =0, \
\end{eqnarray}
such that the redefined fluctuations $\chi(\vec{x},t) =
a^{3/2}\,\, e^{{1\over 2} \int \Gamma dt} \,\delta\phi(\vec{x},t)$
is can be written as a Fourier expansion
\begin{equation}
\chi(\vec{x},t) = \frac{1}{(2\pi)^{3/2}} \int d^3k \left[ a_k
\chi_k(\vec{x},t) + a^{\dagger}_k \chi^*_k(\vec{x},t)\right],
\end{equation}
where $a_k$ and $a^{\dagger}_k$ are the annihilation and creation
operators that comply with the commutation rules:
\begin{equation}
\left[a_k,a^{\dagger}_{k'} \right] = \delta^{(3)}\left(\vec{k}-
\vec{k'}\right), \qquad \left[a^{\dagger}_k,a^{\dagger}_{k'}
\right]=\left[a_k,a_{k'} \right]=0.
\end{equation}

The dynamics of the time dependent modes $\chi_k(t)$ being given
by
\begin{equation}\label{mod}
\ddot\chi_k + \left\{ \frac{k^2}{a^2} - \left[\frac{9}{4}
(H+\Gamma/3)^2 + 3\left((1-2\xi)\dot{H} + \dot{\Gamma}/3\right) -
\left[24\xi H^2 + V''[\phi(t)]\right]\right] \right\} \chi_k(t)=0.
\end{equation}
During inflation the infrared (IR) sector includes long wavelength
modes with $k<k_0$ which are unstable.

\section{Radiation after inflation}

Since the decay width of the produced particles grows with time,
When the inflaton approaches the minimum of the potential, there
is no oscillation of $\phi$ around the minimum energetic
configuration due to at the end of fresh inflation dissipation is
dominant with respect to rate of expansion of the universe:
($\Gamma \gg H$). Hence, the reheating period avoids fresh
inflation). At the end of fresh inflation one would expect that
the universe becomes radiation dominated.

\begin{figure*}
\includegraphics[height=15cm]{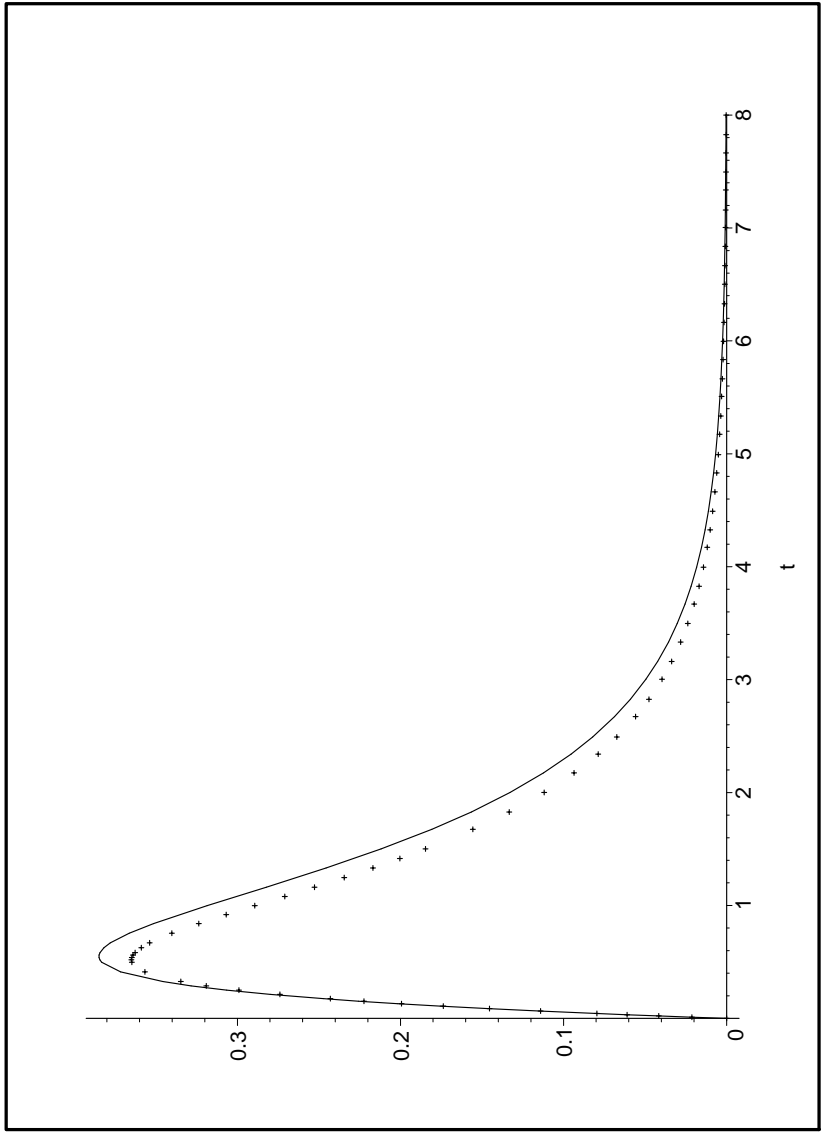}\caption{\label{fig1} Temporal evolution of the cosmological modes $\xi_k(t)/C_1$ for different values of parameters
${k^2 t^2_0 \over 2 M_p} = (0.00001;0.1)$, plotted respectively
with line and dots. }
\end{figure*}

To examine this possibility we consider the case with
$F\simeq4/3$, such that the inflaton field is coupled to gravity
with $\xi=1/4$. This will be possible at the end of fresh
inflation when $V(\phi) \ll \rho_r$ and $\dot\rho_r + 4 F\rho_r
\simeq 0$. Hence the equation (\ref{7}) holds
\begin{equation}
F\simeq \frac{\frac{4}{3} \rho_r}{\rho_r + V} \simeq 4/3.
\end{equation}
For a number of degrees of freedom for relativistic particles of
the order of $g_{eff} \simeq 10^2$ one obtains that the
temperature reaches values like
\begin{equation}
\theta_{Rad}\equiv\left.\theta(t)\right| \simeq 2.5 \times 10^{-5}
\,\, {\rm M_p},
\end{equation}
which is a value very close to the GUT-temperature value and
corresponds to a radiation energy density $\rho^{(0)}_{Rad} \simeq
1.28 \times 10^{-18}\,{\rm M_p}^4$. The effective friction
parameter at this moment will be $\Gamma_{Rad} \simeq {g^4_{eff}
\over 192 \pi} \theta_{Rad}$, which can take the value
\begin{equation}
\Gamma_{Rad} \simeq 4{\rm M_p}.
\end{equation}
After it, the radiation energy density begins to decrease as
$\rho_r(t) = \rho^{(0)}_{Rad} (t/t_0)^{-2}$ and so that
$\dot\rho_r <0$ decreases in a such manner that the temperature
decreases as $\theta_r(t)=\theta_{Rad} (t/t_0)^{-1/2} \sim
a^{-1}(t)$, which is that one expects in a radiation dominated
universe. At this epoch the effective equation of motion
(\ref{mod}) for the modes is
\begin{equation}
\ddot\chi_k + \left\{ \frac{k^2 t_0}{t} - M^2_p \right\}
\chi_k(t)=0,
\end{equation}
when we have made use of the fact that $\Gamma^2_R \gg
\left.(H^2,\dot H, \dot \Gamma, V'')\right|_{t=t_R}$. The general
solution for this equation is
\begin{equation}\label{sol}
\chi_k(t) = C_1 \, \textsc{M}\left[\frac{k^2 t^2_0}{2 M_p},
\frac{1}{2}, 2 M_p t\right] + C_2 \, \textsc{W}\left[\frac{k^2
t^2_0}{2 M_p}, \frac{1}{2}, 2 M_p t\right],
\end{equation}
where ($ \textsc{M}, \textsc{W}$) are the Whittaker functions. In
the UV sector of the spectrum the modes are stable. However, for
$k < M_p \left({t_0\over t}\right)$ the this modes are unstable
and the dominant part of the solution (\ref{sol}) is
\begin{equation}\label{sol1}
\chi_k(t) \simeq C_1 \, \textsc{M}\left[\frac{k^2 t^2_0}{2 M_p},
\frac{1}{2}, 2 M_p t\right] .
\end{equation}
However, notice that the modes $\xi_k(t) = a^{-3/2} e^{-\int
\Gamma_R t}$ increase at early radiation times, but after it decay
to zero with time without crossing the Hubble horizon. It can be
seen in the figure (\ref{fig1}) for the values ${k^2 t^2_0 \over 2
M_p} = (0.00001;0.1)$.

\section{Final Comments}

Reheating is not a minor phase at the end of standard inflation.
Standard inflation may cause the monopole and domain wall
nonthermal symmetry restoration\cite{KLS}, which could give a very
inhomogeneous universe that disagree with observation. This last
possibility is actually rather difficult in simple models of
reheating with only two fields\cite{dv}. However, this situation
can be solved in the case of multiple fields, relevant for GUT
models\cite{bas}. This is the case here studied, where the number
of created particles can be justified by means of the
self-interaction of the inflaton. As in the case of warm
inflation\cite{berera,rombe,nozari}, the main difference between
fresh inflation and standard inflation resides in that here there
is not oscillation of the inflaton field around the minimum of the
potential, due to dissipation is too large at the end of
inflation. Here, particles creation becomes during inflation,
beginning it at zero temperature.

In this work we have examined the natural exit from a fresh
inflationary scenario to a radiation dominated universe when the
inflaton field is coupled to gravity. In the particular case where
the coupling constant is $\xi=1/4$ it is possible to see that this
transition is very natural at the end of inflation in such that
manner that the parameter $F=4/3$ when the inflaton field reaches
its minimum energetic configuration. At this moment all the modes
of the inflaton field fluctuations are stable and the background
temperature is close to the GUT one. However only oscillate the
modes with wavelength below the Planckian scale. The modes which
are in the classical limit are all stable but they increase to
thereafter go down to zero [see figure (\ref{fig1})]. This is
because at this time the inflaton field is strongly coupled to
gravity and the modes decay very faster than during inflation due
to they are super-damped by friction.

\section*{Acknowledgements}

\noindent M. B. acknowledge CONICET and UNMdP (Argentina) for
financial support.

\bigskip

\begin{thebibliography}{99}
\bibitem{Guth}
A. Starobinsky, Phys. Lett. {\bf B91}:99 (1980).\\
A. Guth, Phys. Rev. {\bf D23}:347 (1981).
\bibitem{Al} A. Albrecht and P. J. Steinhardt, Phys. Rev. Lett. {\bf 48}:1220 (1982);\\
A. Linde, Phys. Lett. {\bf B116}:335 (1982).
\bibitem{smoot} R. L. Smoot {\em et al.}, Astrophys. J. {\bf 396}:L1 (1992).
\bibitem{fresh} M. Bellini, Phys. Rev. {\bf D63}: 123510(2001); \\
M. Bellini, Phys. Rev. {\bf D64}: 123508(2001); \\
M. Bellini, Phys.Rev. {\bf D67}: 027303(2003).
\bibitem{berera} A. Berera, Phys. Rev. Lett. {\bf 75}: 3218(1995);
\\
A. Berera, M. Gleiser, R. O. Ramos, Phys. Rev. Lett. {\bf 83}:
264(1999).
\bibitem{coupling} M. Bellini, Gen. Rel. Grav. {\bf 34}:
1953(2002).
\bibitem{14} H. P. de Oliveira and R. O. Ramos, Phys. Rev. {\bf D57}: 741(1998).
\bibitem{17} A. Albrech, P. J. Steinhardt and M. S. Turner, Phys. Rev. Lett. {\bf 48}: 1437(1982).
\bibitem{KLS} L. Kofman, A. Linde, and A. A. Starobinsky, Phys.
Rev. Lett. {\bf 76}: 1011(1996).
\bibitem{dv} D. Boyanovsky, H. J. de Vega, R. Holman and J. F. J. Salgado, Phys. Rev. {\bf D54}:
7570(1996).
\bibitem{bas} B. A. Basset, F. Tamburini, Phys. Rev. Lett. {\bf 81}: 2630(1998).
\bibitem{rombe} J. M. Romero, M. Bellini, Nuovo Cim. {\bf B124}:
861(2009).
\bibitem{nozari} K. Nozari, M. Shoukrani, Astrophys.Space Sci.
{\bf 339}: 111(2012).
\end{thebibliography}
\end{document}